
\magnification=\magstep1
\baselineskip=14pt

\newcount\ftnumber
\def\ft#1{\global\advance\ftnumber by 1
          {\baselineskip=12pt \footnote{$^{\the\ftnumber}$}{#1 \par}}}
\def\ni{\noindent}

\def\fr#1/#2{{\textstyle{#1\over#2}}} 

\def\>{\rangle}
\def\<{\langle}
\def\k#1{|#1\>}
\def\b#1{\<#1|}
\def\tr{{\rm Tr}}
\def\c{\fr1/{\sqrt2}}

\def\ra{\rightarrow}
\font\title = cmbx10 scaled 1440        

\newcount\eqnumber
\def\eq(#1){
    \ifx\DRAFT\undefined\def\DRAFT{0}\fi	
    \global\advance\eqnumber by 1%
    \expandafter\xdef\csname !#1\endcsname{\the\eqnumber}%
    \ifnum\number\DRAFT>0%
	\setbox0=\hbox{#1}%
	\wd0=0pt%
	\eqno({\offinterlineskip
	  \vtop{\hbox{\the\eqnumber}\vskip1.5pt\box0}})%
    \else%
	\eqno(\the\eqnumber)%
    \fi%
}
\def\(#1){(\csname !#1\endcsname)}

\centerline{\title Whose Knowledge?}  \bigskip \centerline{N. David
Mermin} \centerline{Laboratory of Atomic and Solid State Physics}
\centerline{Cornell University, Ithaca, NY 14853-2501} \bigskip
{\narrower \narrower \baselineskip = 12 pt

\noindent  (To appear in {\it Quantum (Un)speakables: Essays in
Commemoration of John S. Bell\/}, eds. Reinhold Bertlmann and Anton
Zeilinger, Springer Verlag, late 2001.)

\vskip 30pt

Sir Rudolph Peierls, in a reply to John Bell's last critique of the
state of our understanding of quantum mechanics, maintained that it
is easy to give an acceptable account of the physical significance of
the quantum theory.  The key is to recognize that all the density
matrix characterizing a physical system ever represents is {\it
knowledge\/} about that system.  In answer to Bell's implicit rejoinder
``{\it Whose\/} knowledge?'' Peierls offered two simple consistency
conditions that must be satisfied by density matrices that convey the
knowledge different people might have about one and the same physical
system: their density matrices must commute and must have a non-zero
product.  I describe a simple counterexample to his first condition,
but show that his second condition, which holds trivially if the first
does, continues to be valid in its absence.  It is an open question
whether any other conditions must be imposed.

}\bigskip 

I met John Bell two and a half times.  The two times, both
memorable for me, were in what turned out to be in the last year of
his life.  The first was at a summer school in Erice in August, 1989,
where he delivered what came close to being the most spell-binding
lecture I have ever heard.  (The only competitors are Richard
Feynman's 1965 Messenger lectures at Cornell.)  The text was
subsequently published in a much quoted article in {\it Physics
World\/}, ``Against Measurement'' [1].  The article conveys his
brilliance and his wit, but not, of course, the music of his voice and
the fire of his beard and hair.

The other moments I remember from our first meeting are Bell's polite
but evidently skeptical response to my telling him that I had long ago
written a book on relativity [2] that took the view he advocated
several years later in ``How to teach special relativity'' [3].  Later
I was pleased to get a letter saying that he had looked up my book and
agreed that I did indeed do it right.  

And I remember a reception at which there suddenly rang out over the
noise of the crowd a cry of ``Don't be a sissy!''  It was John Bell,
encouraging a younger physicist not to let the scope of his speculative
investigations be overly constrained by the wisdom of his elders.

Our second meeting was for a week in Amherst in the summer of 1990.
George Greenstein and Arthur Zajonc had invited about a dozen
people, among them John and Mary Bell, Kurt Gottfried, Viki
Weisskopf, Phillip Pearle, Tony Leggett, Danny Greenberger, Mike
Horne, Anton Zeilinger, and Abner Shimony, to spend a week in a
fraternity house chatting about the nature of quantum mechanics.  No
prepared talks, no schedule, no proceedings.  Just wonderful
conversations.

Shortly before the meeting I had written John --- this was at the very
end of the epoch in which one communicated at a leisurely pace on
paper --- to tell him about a simplified version I had constructed of
a little known argument of Greenberger, Horne, and Zeilinger.  I got
back a wonderful reply, which I have searched for in my files
unsuccessfully ever since he died.  He said that the argument ``filled
me with admiration'', and he wrote down a relevant three-particle Bell
inequality, which inspired me to find a stronger one [4].  (So somewhere,
lost in my filing cabinet, is proof that the very first Bell
inequality for ``Bell's Theorem without Inequalities'' was Bell's Bell
inequality.)  There were lots of entertaining discussions and
arguments in Amherst.  It was far and away the finest conference I
have ever been to.

I also have evidence that over a third of a century ago we had a half
meeting.  The evidence is in the form of a conference photograph in my
office taken in Birmingham England in the summer of 1967 at a meeting
to celebrate the 60th birthday of Rudi Peierls.  Amongst 100
white-shirted necktied gentlemen is a figure in the back row toward
the left in a plaid shirt without a tie.  He sports one of the only
two beards in the photograph and is identified only as Bell; on the
right in the rear is a more conventional looking youth identified as
Mermin.  

This Bell looks like an early version of the John Bell I later knew,
and since the even younger John Bell spent a year in Birmingham in
1953-4 (where I spent two years in 1961-3), it seems clear that the Bell
in the photo was indeed John.  So I could have met John Bell almost a
quarter of a century before I actually did meet him.  It is probable
that I was even introduced to him at that time, quite oblivious of the
fact that I was meeting the discoverer of what would come to be called
Bell's theorem, just as it was quietly tiptoeing onto the intellectual
scene.

Our virtual meeting through Peierls is relevant to my contribution to
this volume, because what I have to say was inspired by another, more
recent, link between Bell and Peierls.  After John's Erice lecture
appeared in {\it Physics World\/}, but too late, alas, for him to read
and reply to it, there appeared something between a commentary and a
rebuttal by Peierls [5], by then, I calculate, a sprightly 84, which
raised for me a puzzle which to this day I have not been able to
answer to my satisfaction.

Peierls agreed with Bell that we should have ``a clearly formulated
presentation of the physical significance of the [quantum] theory
without relying on il-defined concepts'' and that ``I do
not know of any textbook which explains these matters to my
satisfaction.''  On the other hand, he added, ``I do not agree with
John Bell that these problems are very difficult.  I think it is easy
to give an acceptable account.''  According to Peierls all it boils
down to is this:
\medskip
{\narrower\narrower
     \noindent In my view the most fundamental statement of quantum
     mechanics is that the wavefunction, or, more generally the
     density matrix, represents our {\it knowledge\/} of the system we
     are trying to describe.

}\medskip
\noindent Interestingly, {\it knowledge\/} is not on Bell's now famous list of
\medskip

{\narrower\narrower \noindent words which, however legitimate and necessary in
application, have no place in a {\it formulation\/} with any
pretension to physical precision.

}

\medskip\noindent But ``information'' is on the proscribed list, the
charge against it being: 
\medskip
{\narrower\narrower\noindent{\it Information? Whose\/} information?
Information about {\it what?}  

}\medskip\noindent I believe we can take ``information'' and ``knowledge'' to
be synonymous in this context.

Until quite recently I was entirely on Bell's side on the matter of
knowledge-informa\-tion.  But then I fell into bad company.  I started
hanging out with the quantum computation crowd, for many of whom
quantum mechanics is self-evidently and unproblematically all about
information.  I digress to remark that among the tragedies of John
Bell's early death is his missing the quantum computation revolution
of the 1990's.  It has been widely regretted that death deprived us of
Einstein's reaction to Bell's theorem; another great loss is never to
know what Bell would have had to say about Peter Shor's factoring
algorithm or Lov Grover's search algorithm, and the novel view of
quantum mechanics that emerges when you take the theory seriously as a
source of impossibly fast algorithms for the processing of
knowledge-informa\-tion.

Partly from my associations with quantum computer scientists and
partly from endless debates with constructivist sociologists of
science, I have come to feel that ``Information about {\it what?\/}''
is a fundamentally metaphysical question that ought not to distract
tough-minded physicists.  There is no way to settle a dispute
over whether the information is about something objective, or is
merely information about other information.  Ultimately it is a matter
of taste, and, like many matters of taste, capable of arousing strong
emotions, but in the end not really very interesting.

On the other hand ``{\it Whose\/} information?'' raises serious
practical questions that simply have to be addressed.  For while
information, in the form of sense perceptions, is all any of us have
direct access to, we have developed powerful ways to exchange with
each other some of the content of our own private information.  It is
entirely legitimate and unambiguous to raise the question of what
constraints, if any my possession of my knowledge imposes on the
knowledge that you can possess, and vice versa.

Peierls was surely aware of Bell's double-barreled shot at
information-knowledge and immediately after declaring that it's all
about knowledge, he promised to ``return later to the question `whose
knowledge'$\,$''.  I infer that like the very recent me, he did not
find it necessary to deal with the question ``knowledge about what?''
But towards the end of his article he dealt briskly and efficiently
with ``whose knowledge?'':
\medskip
     {\narrower\narrower \noindent [Density matrices] may differ,
     as the nature and amount of knowledge may differ.  People may
     have observed the system by different methods, with more or less
     accuracy; they may have seen part of the results of another
     physicist.  However, there are limitations to the extent to which
     their knowledge may differ. This is imposed by the uncertainty
     principle. For example if one observer has knowledge of $S_z$ of
     our Stern-Gerlach atom, another may not know $S_x$, since
     measurement of $S_x$ would have destroyed the other person's
     knowledge of $S_z$, and vice versa.  This limitation can be
     compactly and conveniently expressed by the condition that the
     density matrices used by the two observers must commute with each
     other.

}\medskip\noindent In a chapter in {\it More Surprises in Theoretical
Physics\/} to which he referred the reader for further details,
Peierls added a second condition, which is easy to understand if the
density matrices do indeed commute [6]:
\medskip

{\narrower\narrower \noindent At the same time, the two
observers should not contradict each other.  This means the product of
the two density matrices should not be zero.

}
\medskip

Peierls' rationale for his first condition has a highly plausible ring
to it.  After all, if Alice {\it knows\/} that a spin-$\fr1/2$
particle is in an eigenstate of $S_z$, then absolutely no information
can be available to anybody about any components of its spin
orthogonal to the $z$-direction.  A measurement of any spin component
in the $x$-$y$ plane is as likely to come out up as down.
Mathematically this requires that nobody's density matrix can have
off-diagonal elements in the $S_z$ representation --- i.e. everybody's
density matrix must commute with the pure-state density matrix of
Alice.

Things become a little less clear if Alice's knowledge is not
characterized by a pure-state density matrix, and considerably less
clear if we are dealing with more than a two-state system.  For some
time I struggled to find a general way to understand Peierls' first
criterion.  I stopped struggling when Chris Fuchs pointed out to me
that it was clearly incorrect, even for two-state systems and even
when Alice's density matrix described a pure state.  I give here a
slightly embellished version of Fuchs' simple counterexample, which
makes explicit how a disparity in knowledge can arise that leads Alice
and Bob to describe one and the same physical system with
non-commuting density matrices.

Consider a pair of qubits (spins-$\fr1/2$, if you prefer, 0 being
$\uparrow$ along {\bf z} and 1 being $\downarrow$) in the state $$
\k\Psi = \cos\theta\k0\k0 + \sin\theta\k1\k1. \eq(2qubit)$$ with $0 <
\theta < \pi/4$. After the state has been prepared the qubits cease to
interact.  To emphasize the absence of subsequent interaction the
qubit on the right (hereafter called the right qubit) can be carried
far away from the other (hereafter called the left qubit).  Anybody
who knows how the two-qubit state was prepared --- and, in particular,
Alice and Bob, who both know --- will agree that although the
two-qubit system is in a pure state, each separate qubit, and in
particular the left qubit, has no pure state of its own but is
characterized, at the most fundamental level, by the mixed-state
density matrix $$\rho = \tr_r\k\Psi\b\Psi = \cos^2\theta\,\k0\b0\ +\
\sin^2\theta\,\k1\b1,\eq(rho)$$
where $\tr_r$ is the partial trace over the degrees of freedom of the
right qubit.

Suppose Alice now goes to the right qubit and secretly measures it in
the computational basis.  She does not report to Bob the result of her
measurement or even whether she has measured it at all.  Since the
right qubit is far away and does not interact with the left qubit and
since Alice communicates no information whatever to Bob, regardless of
what she has or has not done to the right qubit, Bob must surely
continue to describe the left qubit with the same density matrix
\(rho). Alice, on the other hand, knowing the strong correlations
present in the state \(2qubit), is able after measuring the right
qubit to update the density matrix she uses to express her knowledge
about the qubit on the left to one of the pure-state density matrices
$\k0\b0$ or $\k1\b1$, depending on whether the result of her
measurement on the right is 0 or 1.  Since either of these commutes
with Bob's density matrix \(rho), Peierls' first condition is
confirmed.

Suppose, however, that prior to making her measurement Alice secretly
applies to the right qubit the unitary Hadmard transformation, $$ \k0
\ra \c(\k0+\k1),\ \ \ \k1 \ra \c(\k0-\k1).\eq(hadamard)$$ This
transforms the two-qubit state into $$\eqalign{\k\Phi &= \c\bigl(
\cos\theta\,\k0(\k0+\k1) + \sin\theta\,\k1(\k0-\k1)\bigr)\cr &=
\c\bigl((\cos\theta\,\k0 + \sin\theta\,\k1)\k0 + (\cos\theta\,\k0-
\sin\theta\,\k1)\k1\bigr).}\eq(posthadamard)$$

Now, depending on whether Alice gets 0 or 1 from her measurement of
the right qubit, she will assign the left qubit one or the other of
the pure-state density matrices $$\eqalign{\rho_a &=
\bigl(\cos\theta\,\k0 \pm \sin\theta\,\k1\bigr) \bigl(\cos\theta\,\b0
\pm \sin\theta\,\b1\bigr) \cr &= \cos^2\theta\k0\b0 +
\sin^2\theta\k1\b1 \pm\cos\theta\sin\theta \bigl(\k0\b1 +
\k1\b0\bigr).}\eq(otherrhoa)$$ But Bob, who does not know what, if
anything Alice did to the right qubit, must continue to use \(rho) for
his density matrix $\rho_b$ describing the left qubit.  We therefore
have $$[\rho_b, \rho_a] =
\pm\fr1/4\sin4\theta\bigl(\k0\b1-\k1\b0).\eq(comm)$$ This commutator
is non-zero for $0 < \theta < \pi/4$ regardless of the result of
Alice's measurement.  Peierls' criterion fails to hold.

Having lived and worked for two years in Peierls' Birmingham
department, where I came to admire him immensely, I am adding to my
list of reactions I wish death had not snatched from us Peierls'
response to Fuchs' counterexample.  All I know for sure, is that he
would not have attempted to escape by maintaining that Alice's action
on the right qubit had somehow altered the objective condition of the
left qubit.  He would have agreed that whatever she did to and learned
about the qubit on the right altered only her {\it knowledge\/} of the
faraway qubit on the left.  Since Alice communicates none of this
updated knowledge to Bob, there is no way for him to update his
density matrix as a result of her actions.  But since she now {\it
knows\/} that the density matrix for the left qubit is a particular
pure state density matrix, she is surely entitled to use it as the
appropriate expression of her current knowledge.

Here is what seems to go wrong with Peierls' justification for his
first condition.  Depending on the result of her measurement on the
right qubit, Alice's knowledge of the left qubit is indeed
encapsulated in one of the two pure states $$\k\psi_+ =
\cos\theta\,\k0 + \sin\theta\,\k1\eq(pure+)$$ or $$\k\psi_- =
\cos\theta\,\k0 - \sin\theta\,\k1.\eq(pure-)$$ In spin language she
therefore either knows that the left qubit is in an eigenstate of the
spin along the direction {\bf n}$_+$ = cos$2\theta\,${\bf z} +
sin$2\theta\,${\bf x} or she knows that it is in an eigenstate of the
spin along the direction {\bf n}$_-$ = cos$2\theta\,${\bf z} -
sin$2\theta\,${\bf x}.  As Peierls noted, it is inconsistent with
Alice's knowledge for Bob to know anything about the spin along any
direction orthogonal to {\bf n}$_+$ in the former case, or orthogonal
to {\bf n}$_-$ in the latter case.

But Bob does not know which case is the case.  Indeed, he does not
even know whether Alice knows which case is the case and therefore (to
get slightly metaphysical) whether there {\it is} a case which is the
case.  Should Alice happen to know that the state of the left qubit is
$\k\psi_+$, then it is not incompatible with Alice's knowledge for Bob
to be able to make more than a completely random guess about the
result of measuring a spin component in the plane perpendicular to
${\bf n}_-$, and vice-versa.  The only direction along which Bob
cannot improve on a random guess, regardless of whether $\k\psi_+$ or
$\k\psi_-$ encapsulates Alice's knowledge of the left qubit, is the
direction {\bf y} orthogonal to both ${\bf n}_+$ and ${\bf n_-}$.  And
Bob's density matrix \(rho) does indeed give equal probabilities for
his finding either value of $S_y$.  That this state of affairs holds
whatever unitary transformation Alice applies to the right qubit
before she measures it in the computational basis, is shown in the
Appendix below.  So Bob gets away with violating Peierls' constraint
on what he can know because he does not know what knowledge Alice
knows he cannot have, and therefore must continue to use a density
matrix that embodies knowledge that it {\it might\/} be possible for
him to have.

I do not believe that this counterexample is contrived or artificial.
The initial density matrix \(rho) that both Alice and Bob assign to
the left qubit is required by basic principles of quantum mechanics
and the fact that what each of them knows about the two-qubit system
is given by the pure state \(2qubit).  Alice's subsequent acquisition
of information enabling her to refine her density matrix down to one
or the other of the two choices of pure-state density matrix
\(otherrhoa) cannot alter Bob's assignment of density matrix, because
the actions she takes to make the refinement are taken far from the
left qubit.  There is therefore no material alteration of the left
qubit that Bob is, in some way, willfully ignoring.  Nor can there be
any change in Bob's {\it knowledge\/} of the left qubit, since neither
the nature of Alice's actions nor the outcomes of those actions are
communicated to Bob.

So to appropriate a remark John Bell once made about the
Einstein-Podolsky-Rosen argument [7], Peierls' first condition
``doesn't work.  The reasonable thing just doesn't work.''  We are
left with the question of what constraints of mutual consistency, if
any, one {\it can\/} impose on a collection of density matrices that
encapsulate the knowledge available to Alice, Bob, Charles, Doris,
Edward,$\,\ldots$ about one and the same physical system.
I do not know the answer.  What I can show, at a minimum, is that
Peierls' second condition remains a valid one, even though he
justified it only under the assumption that his primary condition
held.

Peierls' second condition is the concise mathematical expression of
the fact that whatever other mutual consistency conditions one might
argue for or against, there is one rock-bottom requirement.  The
different probabilities their density matrices $\rho_a$ and $\rho_b$
enable Alice and Bob to assign to the outcome of any subsequent
measurement made on one and the same individual physical system cannot
contradict one another.  This is a very weak restriction, since
neither of their probability assignments can be refuted by the
subsequent occurrence or nonoccurence of any measurement outcome they
both assign a probability that is neither 1 nor 0.  The knowledge one
of them has can contradict the knowledge possessed by the other only
if one of them knows that something will certainly happen, while the
other knows that it certainly will not.  So a minimalist consistency
condition is that there can be no outcome of any measurement to which
one of them assigns a probability 1 and the other assigns a
probability 0.  This consistency condition turns out to be equivalent
to Peierls' condition $$\rho_a\rho_b \neq 0.  \eq(prod)$$

To see this note first that if \(prod) fails to hold, so that
$\rho_a\rho_b = 0$, then since density matrices are hermitian we also
have $\rho_b\rho_a = 0$.  So $\rho_a$ and $\rho_b$ commute, and there
is a basis of joint eigenstates.  Since the vanishing of
$\rho_a\rho_b$ requires at least one of $\rho_a$ and $\rho_b$ to have
zero eigenvalue for each eigenstate, we can resolve the identity (not
necessarily uniquely) into the sum of two projections, $$ 1 = P_a +
P_b, \eq(resolve)$$ where $P_a$ and $P_b$ project onto subspaces of
eigenstates of $\rho_a$ and $\rho_b$ with eigenvalue zero.  The
outcomes of any measurement that discriminates between these two
orthogonal subspaces will be assigned probabilities 1 or 0 by Alice,
and 0 or 1 by Bob.  Whatever Alice says must happen Bob will say
cannot happen, and vice versa.

The converse is slightly more subtle.  Suppose there is some
particular measurement with an outcome that Alice gives probability 0
and Bob gives probability 1.  This means there is some hermitian
operator\ft{If you take an old-fashioned view of what constitutes a
measurement it suffices to take the hermitian operator to be a
projection operator --- i.e. every eigenvalue $m$ is either 0 or 1.
If you prefer to think of measurements in terms of
positive-operator-valued measures (POVMs), there is no reason to make
that restriction.  The argument that follows works for either reading
of ``any measurement''.  It thereby demonstrates that the existence of
any such generalized measurement implies the existence of an ordinary
von Neumann measurement.} $M$ with eigenvalues $m$ in the range $$0
\leq m \leq 1\eq(spectrum)$$ with $$\tr\rho_a M = 0,\ \ \ \tr\rho_b M
= 1.\eq(contradict)$$ If the eigenstates of $M$ are $\k{\psi_j}$ with
eigenvalues $m_j$ then expanding \(contradict) in the basis of those
eigenstates gives $$\sum_j\b{\psi_j}\rho_a\k{\psi_j}m_j = 0,
\eq(rhoam0)$$ and $$\sum_j\b{\psi_j}\rho_b\k{\psi_j}m_j = 1,
\eq(rhobm1)$$ or, since $\tr\rho_b = 1$,
$$\sum_j\b{\psi_j}\rho_b\k{\psi_j}(1-m_j) = 0. \eq(rhobm0)$$ Since the
diagonal matrix elements of $\rho_a$ and $\rho_b$ are non-negative and
since the $m_j$ satisfy \(spectrum), we conclude from \(rhoam0) and
\(rhobm0) that $$\b{\psi_j}\rho_a\k{\psi_j} = 0,\ \ \ m_j\neq
0,\eq(rhoa0)$$ $$\b{\psi_j}\rho_b\k{\psi_j} = 0,\ \ \ m_j\neq
1.\eq(rhob0)$$

But if $$\b\psi\rho\k\psi = 0\eq(rho0)$$ for a density matrix $\rho$,
then we must have\ft{For a density matrix $\rho$ has a hermitian
square root, $\sqrt\rho$. Since \(rho0) requires the norm of
$\sqrt\rho\k\psi$ to vanish, so must $\sqrt\rho\k\psi$ itself. So
$\rho\k\psi = \sqrt\rho\sqrt\rho\k\psi = 0.$} $$\rho\k\psi =
0.\eq(eigenstate)$$ Therefore \(rhoa0) and \(rhob0) require every
eigenstate $\k{\psi_k}$ of $M$ to be an eigenstate with zero eigenvalue
of either $\rho_a$ or $\rho_b$.  It follows that
$$\b{\psi_i}\rho_a\rho_b\k{\psi_j} =
\sum_k\b{\psi_i}\rho_a\k{\psi_k}\b{\psi_k}\rho_b\k{\psi_j} = 0
\eq(rhorho0)$$ for arbitrary $i$ and $j$.  Since the $\k{\psi_i}$ are a
complete set, we then have $$\rho_a\rho_b = 0.\eq(rhorho)$$ 
The existence of any measurement whatever with an outcome that Alice's
density matrix requires and Bob's forbids (or vice versa) means that
the product of their density matrices must vanish.

So Bob's knowledge can contradict Alice's if and only if their density
matrices violate Peierls' second condition \(prod).  This minimalist
constraint \(prod) clearly generalizes to any number of density matrices
$\rho_a$, $\rho_b$, $\rho_c\ldots$ used by Alice, Bob, Carol,$\ldots$
to encapsulate what each of them knows about one and the same system.
No two of them can assign probabilities of 1 and 0 to the outcome of
any measurement, and therefore all possible pairs of density matrices
must have a non-zero product.

I do not know if one can impose any conditions beyond requiring all
pairs of density matrices to satisfy the minimalist condition \(prod).
I have the feeling that if quantum mechanics is really about knowledge
and only knowledge, then there ought to be further elementary
constraints on the possible density matrices describing one and the
same physical system that are stronger than the very weak second
condition of Peierls, but not as strong as his overly restrictive
first condition.  I wish somebody would tell me what they are, or
provide me with a convincing argument that there are none.  

And I wish it were still possible to talk about these things with John
Bell.

\bigskip

{\sl Acknowledgment.\/} For some time I went around like Coleridge's
Ancient Mariner, cornering people and asking if they could explain to
me Peierls' first criterion.  It was Chris Fuchs who finally pointed
out to me that the criterion obviously fails if Alice knows the pure
state of a qubit but Bob knows only that it is one or the other of two
nonorthogonal states.  A remark by Ben Schumacher turned my attention
to Peierls' second criterion, and I have had a stimulating
correspondence about it with Rudiger Schack.  This work was supported
by the National Science Foundation, Grants PHY9722065 and PHY0098429.

\bigskip
\centerline{{\bf Appendix}}
\bigskip

Every pure spin-$\fr1/2$ state is a spin-up eigenstate along some
direction {\bf n} and the projection operator on such an eigenstate is
$\fr1/2(1 + {\bf n}\cdot\sigma)$.  So regardless of what observable
Alice chooses to measure on the right qubit --- i.e. regardless of
what unitary transformation she applies before measuring it in the
computational basis --- the post-measurement mixed-state density
matrix for the left qubit can be expressed as a probability-weighted
sum of the two possible post-measurement pure state density matrices
known to Alice as a result of her measurement of the right qubit:
$$\rho = {p \over 2}(1 + {\bf n}\cdot\sigma) + {q \over 2}(1 + {\bf
m}\cdot\sigma), \eq(pq)$$ for probabilties $p$ and $q=1-p$.  But the
density matrix for the left qubit is unaltered by anything Alice does
to the right qubit.  It must therefore continue to be the reduced
density matrix \(rho), which can be rewritten as $$\rho = \fr1/2(1 +
\cos2\theta\,{\bf z}\cdot\sigma).\eq(cos2t)$$ Thus $p$, {\bf m}, and
{\bf n} are subject to the constraint that $${\bf z} =
\sec2\theta\,(p{\bf n} + q{\bf m}).\eq(constraint)$$ So the unique
direction orthogonal to both {\bf n} and {\bf m}, along which quantum
mechanics forbids Bob to have any knowledge of the spin regardless of
the outcome of Alice's measurement, is necessarily orthogonal to {\bf
z}. Bob's density matrix \(rho) does indeed afford him no knowledge
whatever of the spin along that one special direction.  \bigskip

\vfil\eject
\centerline{\bf References} \def\ni{\noindent}

\parskip = 10pt

\ni 1. John Bell,``Against Measurement'', Physics World,
August 1990, 33-40.

\ni 2.  N. David Mermin, {\it Space and Time in Special Relativity\/},
McGraw Hill, New York, 1968; reissued by Waveland Press, Prospect
Heights, Illinois, 1989. 

\ni 3.  J. S. Bell, ``How to teach special relativity'', Progress in
Scientific Culture {\bf 1}, No 2, summer 1976; reprinted in
{\it Speakable and unspeakable in quantum mechanics\/}, Cambridge
University Press, Cambridge, 1987, 67-80.

\ni 4.  N. David Mermin, Phys.~Rev.~Lett.~{\bf 65}, 1838-40 (1990).

\ni 5. R. E. Peierls, ``In defense of `measurement'$\,$'', Physics
World, January 1991, 19-21.

\ni 6. Rudolf Peierls, {\it More Surprises in Theoretical Physics\/}, 
Princeton University Press, Princeton, New Jersey, 1991, 11. 

\ni 7. Quoted by Jeremy Bernstein in {\it Quantum Profiles\/},
Princeton University Press, Princeton, New Jersey, 1991, 84.

\bye